\documentclass[superscriptaddress,12pt,preprint,showpacs]{revtex4}
\usepackage{amssymb} \RequirePackage{graphicx}

\begin{document}

\title{Optical Bistability in Nonlinear Optical Coupler with Negative Index Channel}
\author{Natalia\,M.\,Litchinitser}
\email{natashan@eecs.umich.edu} \affiliation{Department of
Electrical Engineering and Computer Science, University of
Michigan, \\
2200 Bonisteel Boulevard, 3113 ERB1, Ann Arbor, Michigan 48109, USA}
\author{Ildar\,R.\,Gabitov}
\email{gabitov@math.arizona.edu}
\affiliation{Department of Mathematics, University of Arizona\\
617 North Santa Rita Avenue, Tucson, AZ 85721, USA}
\affiliation{L.D. Landau Institute for Theoretical Physics, Russian Academy of Sciences\\
2 Kosygin Street, Moscow, 119334, Russian Federation}
\author{Andrei\,I.\,Maimistov}
\email{maimistov@pico.miphi.ru}
\affiliation{Department of Solid State Physics, Moscow Engineering Physics Institute\\
Kashirskoe sh. 31, Moscow, 115409, Russian Federation}
\date{\today}
\begin{abstract}
We discuss a novel kind of nonlinear coupler with one channel filled
with a negative index material (NIM). The opposite directionality of
the phase velocity and the energy flow in the NIM channel
facilitates an effective feedback mechanism that leads to optical
bistability and gap soliton formation.
\end{abstract}


\maketitle

Nonlinear optical couplers have attracted significant attention
owing to their strong potential for all-optical processing
applications, including switching and power-limiting devices.
Transmission properties of a nonlinear coherent directional coupler
were originally studied by Jensen~\cite{1}, who concluded that a
coupler consisting of two channels made of conventional homogeneous
nonlinear materials is not bistable.

\begin{figure}[ht]
\centering
\includegraphics[scale=0.9]{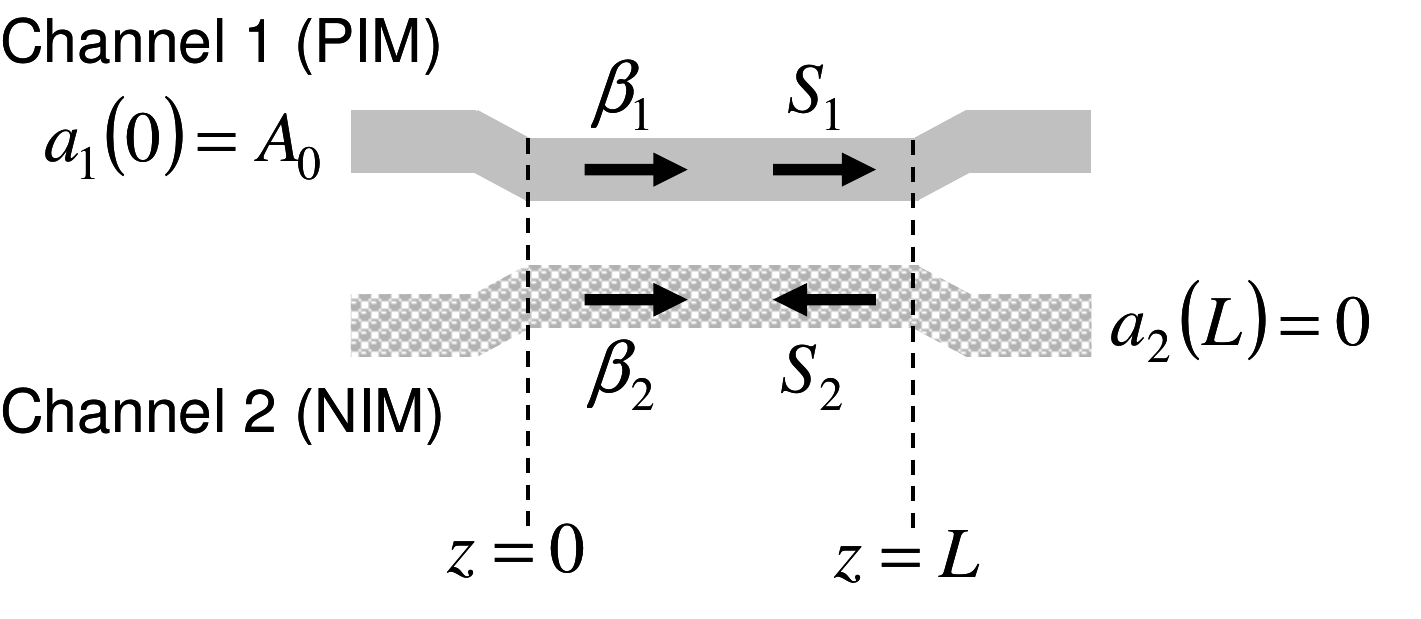}
\caption{\emph{A schematic of a nonlinear PIM-NIM coupler. Light is
initially launched into channel 1 (PIM). A wave vector and a
Poynting vector are parallel in the PIM channel and anti-parallel in
the NIM channel enabling a new backward-coupling
mechanism.}}\label{Fig1}
\end{figure}

Bi- (or multi-)stability is a phenomenon in which the system
exhibits two (or more) steady transmission states for the same input
intensity~\cite{2,3}. Optical bistability has been predicted and
experimentally realized in various settings including a Fabry-Perot
resonator filled with a nonlinear material~\cite{3} and layered
periodic structures~\cite{4}. In this Letter we describe a novel
nonlinear optical coupler structure that utilizes a negative index
metamaterial (NIM)~\cite{5,6} in one of the channels and a
conventional positive index material (PIM) in another channel as
shown in~Fig.~\ref{Fig1}. The linear transmission properties of a
similar optical structure were previously studied by Alu and
Engheta~\cite{7}. We show that such nonlinear coupler (NLC) can be
bistable. Bistability occurs owing to the effective feedback
mechanism enabled by a fundamental property of NIMs -- opposite
directionality of the wave vector and the Poynting vector. Moreover,
our results suggest that the entirely uniform PIM-NIM coupler
structure supports gap solitons -- a feature commonly associated
with periodic structures~\cite{4},~\cite{8}-\cite{14}.

Continuous wave propagation in a nonlinear coupler can be described
by the following system of equations:
\begin{eqnarray}
i\sigma_{1}\frac{\partial a_{1}}{\partial z}+\kappa_{12}a_{2}\exp
(-i\delta z)+\frac{2\pi \omega_{0}}{c}\sqrt{\frac{\mu_{1}(\omega
_{0})}{\epsilon_{1}(\omega_{0})}}\chi_{1}^{(3)}|a_{1}|^{2}a_{1}
&=&0,
\label{eq1a} \\
i\sigma_{2}\frac{\partial a_{2}}{\partial z}+\kappa_{21}a_{1}\exp
(i\delta z)+\frac{2\pi \omega_{0}}{c}\sqrt{\frac{\mu_{2}(\omega
_{0})}{\epsilon_{2}(\omega_{0})}}\chi_{2}^{(3)}|a_{2}|^{2}a_{2}
&=&0,  \label{eq1b}
\end{eqnarray}
where $a_{1}$ and $a_{2}$ are the complex amplitudes of the modes in
the PIM and NIM channels respectively, $\kappa_{12}$ and $\kappa
_{21}$ are the coupling coefficients defined as in Ref.~\cite{1},
$\delta =\beta_{1}-\beta_{2}$ is the mismatch between the
propagation constants in the individual channels, $\beta
_{j}^{2}=(\omega_{0}/c)^{2}\mu_{j}(\omega_{0})\varepsilon
_{j}(\omega_{0})$, ($j=1,2$), $\sigma_{j}$ is the sign of the
 refractive index  $n_{j}=\sqrt{\mu
_{j}(\omega_{0})\varepsilon_{j}(\omega_{0})}$, $\varepsilon_{j}$ and
$\mu_{j}$ are linear frequency-dependent dielectric permittivity and
magnetic permeability, $\omega_{0}$ is carrier frequency, $\chi
_{j}^{(3)}$ is the nonlinear (electric) susceptibility, and $c$ is
the speed of light in a vacuum. In the case of PIM-NIM coupler
$\sigma_{1}$ is positive, while $\sigma_{2}$ is negative.

\begin{figure}[ht]
\centering
\includegraphics[scale=0.9]{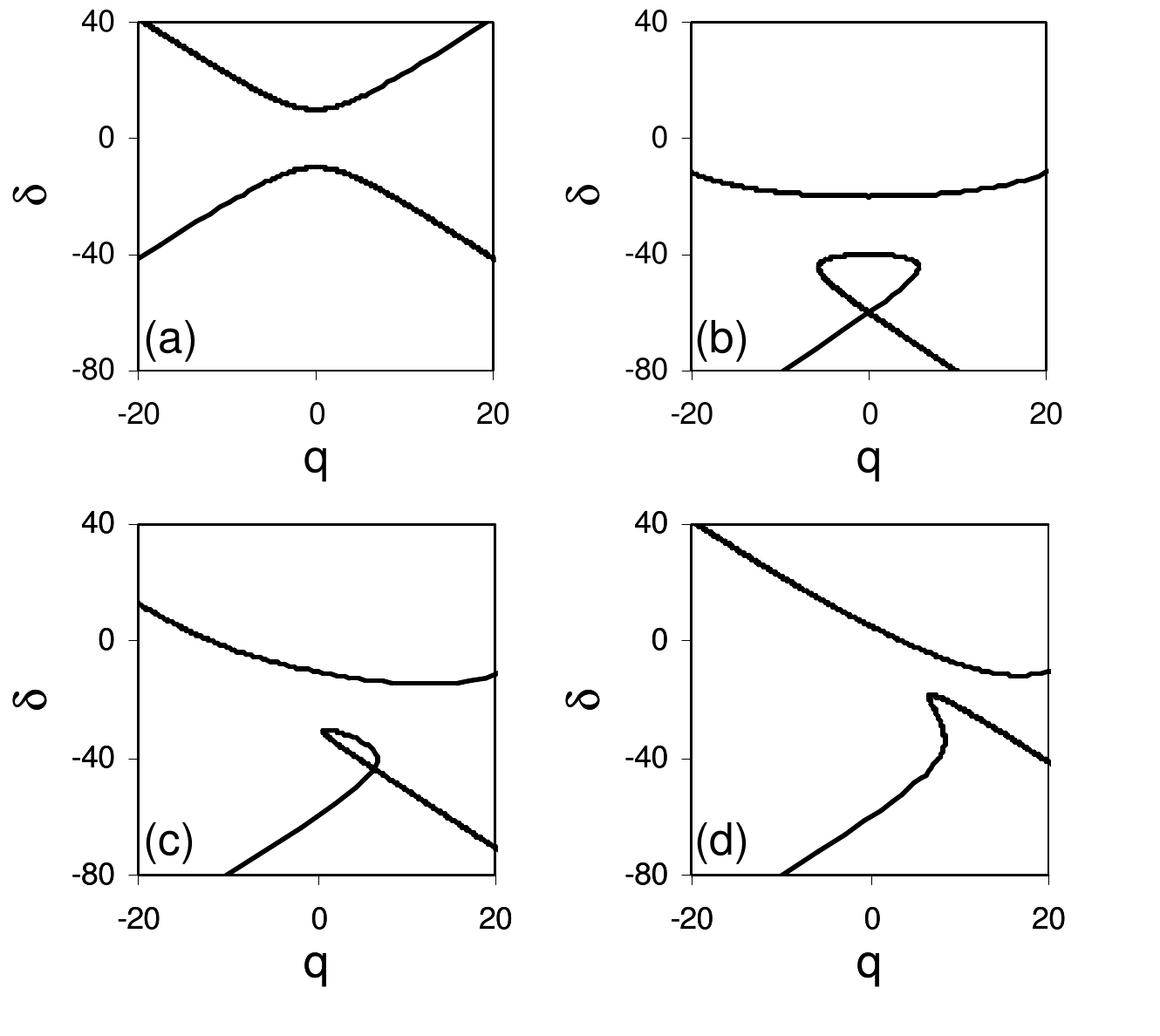}
\caption{\emph{PIM-NIM coupler $\delta -q$ relations:  (a) linear
($\kappa_{12}=\kappa_{21}=5$), (b) two channels with the same
nonlinear susceptibility $\gamma_{1}=\gamma_{2}=\gamma $ ($\kappa
_{12}=\kappa_{21}=5$, $\alpha ^{2}\gamma/\kappa =6$), (c) two
channels with different nonlinear susceptibilities
($\kappa_{12}=\kappa_{21}=5,~\alpha ^{2}\gamma_{1}/\kappa =6,~\alpha
^{2}\gamma_{2}/\kappa =3$), (d) the NIM channel is linear, while the
PIM channel is nonlinear ($\kappa _{12}=\kappa_{21}=5,~\alpha
^{2}\gamma_{1}/\kappa =6,~\alpha ^{2}\gamma_{2}/\kappa
=0$).}}\label{Fig2}
\end{figure}

Assuming the form $a_{1,2}=u_{1,2}\exp (iqz)\exp(\mp i \delta z/2)$
for the solutions of Eqs.~(\ref{eq1a}, \ref{eq1b}) in the linear
regime, we find the following relation between $q$\ and $\delta $
for the PIM-NIM coupler
\begin{equation}
q^{2}=(\delta /2)^{2}-\kappa_{12}\kappa_{21},  \label{eq2}
\end{equation}
which indicates the presence of a bandgap for $|\delta |\leq
2\sqrt{\kappa_{12}\kappa_{21}}$. The photonic bandgap is usually a
feature typical for periodic or distributed feedback (DFB)
structures such as fiber Bragg gratings (FBGs) or thin film
stacks~\cite{7}. Formation of the bandgap in a uniform structure
considered here is one of the unique properties of the PIM-NIM
coupler arising from introduction of the NIM into the nonlinear
coupler. Introducing two parameters $a^{2}=u_{1}^{2}+u_{2}^{2}$ and
$f=u_{2}/u_{1}$, the nonlinear counterpart of the relation~\ref{eq2}
can be written in the form
\begin{eqnarray}
\delta  &=&-\frac{\kappa_{21}+f^{2}\kappa
_{12}}{f}-\frac{a^{2}(\gamma
_{1}+\gamma_{2}f^{2})}{1+f^{2}},  \label{eq3a} \\
q &=&-\frac{\kappa_{21}-f^{2}\kappa_{12}}{2f}-\frac{a^{2}(\gamma
_{2}f^{2}-\gamma_{1})}{2(1+f^{2})}  \label{eq3c}
\end{eqnarray}
where
\[
\gamma_{j}=\frac{2\pi \omega_{0}}{c}\sqrt{\frac{\mu_{j}(\omega
_{0})}{\epsilon_{j}(\omega_{0})}}\chi_{j}^{(3)}
\]
is the nonlinearity coefficient. Figure~\ref{Fig2}  shows linear (a)
and nonlinear $\delta -q$ curves (b)-(d) for the case of
self-focusing Kerr nonlinearity. In Fig.~\ref{Fig2}(b) both PIM and
NIM channels are nonlinear, with the same nonlinearity coefficients.
Beyond a critical power level, the lower branch of the $\delta -q$
curve forms a loop. Both linear and nonlinear $\delta -q$\ curves in
Fig.~\ref{Fig2}(a) and~\ref{Fig2}(b) resemble dispersion relations
found in the case of linear and nonlinear FBG~\cite{8,15}, with the
difference that in the nonlinear case the effect of cross-phase
modulation has been neglected and forward and backward waves are
spatially separated. However, importantly, in the case of PIM-NIM
NLC both channels are made of homogeneous material with no
periodicity or an external feedback mechanism (such as a cavity).
However, the effective feedback mechanism is provided by the
inherent property of the NIMs, that is opposite directionality of
the phase and energy velocities. As shown in Fig.~\ref{Fig1}, while
the propagation vectors of the waves propagating in both NLC
channels point in the same direction assuring a necessary
phase-matching condition, the Poynting vectors corresponding to the
energy flow direction point in opposite directions. As light
propagates in the PIM channel in the forward direction it
continuously couples to the NIM channel where it flows in the
backward direction. Therefore, the PIM-NIM coupler acts as an
effective DFB structure. Figure~\ref{Fig2}(c) corresponds to the
case of different nonlinear coefficients in the two channels, while
Fig.~\ref{Fig2}(d) corresponds to the case of the nonlinear PIM and
linear NIM channel. Figures~\ref{Fig2}(c) and~\ref{Fig2}(d) also
illustrate that there are more degrees of design freedom in the NLC
case in comparison with the DFB structures since in general case the
nonlinear coefficients $\gamma_{1}$ and $\gamma_{2}$ as well as the
coupling coefficients $\kappa_{12}$ and $\kappa_{21}$ may be not
identical and can be properly designed.

Although in general case $\kappa_{12}\neq \kappa_{21}$ and
$\gamma_{1}\neq \gamma_{2}$, in order to illustrate basic new
physical effects associated with the PIM-NIM NLC, in the following
discussion we assume identical linear coupling coefficients $\kappa
_{12}=\kappa_{21}\equiv \kappa $ and nonlinear coefficients $\gamma
_{1}=\gamma_{2}\equiv \gamma $. Then, making the substitution
$a_{1}=A_{1}\exp (i\phi_{1})$ and $a_{2}=A_{2}\exp (i\phi_{2})$,
where $A_{1}$, $A_{2}$, $\phi_{1}$ and $\phi_{2}$ are real functions
of $z$, the equations for $A_{1},$\ $A_{2}~$and $\psi
=\phi_{1}-\phi_{2}+\delta z$ can be written in the form
\begin{eqnarray}
\frac{\partial A_{1}}{\partial z} &=&\kappa A_{2}\sin \psi \,\nonumber \\
\frac{\partial A_{2}}{\partial z} &=&\kappa A_{1}\sin \psi ,  \label{eq4b} \\
\frac{\partial \psi }{\partial z} &=&\kappa \left(
\frac{A_{2}}{A_{1}}+\frac{ A_{1}}{A_2}\right) \cos \psi +\gamma
\left( A_{1}^{2}+A_{2}^{2}\right) +\delta .\nonumber
\end{eqnarray}
From the system of equations~(\ref{eq4b}) the constants of the
motion are given by
\begin{eqnarray}
C &=&P_{1}-P_{2},  \label{eq5} \\
\Gamma  &=&4A_{1}A_{2}\cos \psi +2\kappa ^{-1}A_{1}^{2}\left( \delta
+\gamma A_{2}^{2}\right)   \label{eq5b}
\end{eqnarray}
where $P_{1}=A_{1}^{2}$ and $P_{2}=A_{2}^{2}$. $C$ is defined by the
boundary conditions at $z=0$ and $z=L$, where $L$ is the length of
the coupler. The expressions~(\ref{eq5}) should be compared to that
of conventional PIM-PIM NLC, in which case $C=P_{1}+P_{2}$.

Then, power evolution in channel 1 is described by the equation
\begin{equation}
\left( dP_{1}/dz \right) ^{2}=\kappa (4\kappa +\Gamma \gamma
)P_{1}P_{2}+\Gamma \kappa \delta P_{1}-\kappa ^{2}\Gamma
^{2}/4-(\delta +\gamma P_{2})^{2}P_{1}^{2}. \label{eq8}
\end{equation}

If light initially is launched into channel 1, i.e.
$A_{1}(0)=A_{0}$, $ A_{2}(L)=0$, then $C=A_{1}^{2}(L)$. In the case
of $\delta =0$ and $\Gamma =0 $ Eq.~(\ref{eq8}) reduces to
\begin{equation}
\left( dP_{1}/dz\right) ^{2}=4\kappa ^{2}P_{1}(P_{1}-C)-\gamma
^{2}P_{1}^{2}(P_{1}-C)^{2}.  \label{eq9}
\end{equation}
Introducing new variable $Z=C\kappa \alpha z$ and $y^{2}=\alpha
^{2}P_{1}(P_{1}-C)$, where $\alpha =\gamma /2\kappa $,
Eq.~(\ref{eq9}) can be written in the form
\begin{equation}
(dy/dZ)^{2}=(1-y^{2})(1+k^{2}y^{2}),  \label{eq10}
\end{equation}
where $k=(\alpha C/2)^{-1}$. The solution of Eq.~(\ref{eq10}) is
given by
\begin{equation}
y(Z)=\text{\textrm{sn}}\left( Z-Z_{1};ik\right) ,  \label{eq11}
\end{equation}
where $Z_{1}=C\kappa \alpha L$ as follows from the boundary
condition, \textrm{sn}$\mathrm{(}z^{\prime },k^{\prime }\mathrm{)}$\
is the Jacobi elliptic function~\cite{16}. Using the properties of
Jacobi elliptic functions and Eq.~(\ref{eq11}) the solutions for
$P_{1}$\ and $P_{2}$\ can be written in the form
\begin{eqnarray}
P_{1}(z) &=&C\frac{1+\mathrm{dn}[2\kappa
(z-L)/m;m]}{2\mathrm{dn}[2\kappa
(z-L)/m;m]},  \label{eq12a} \\
P_{2}(z) &=&C\frac{1-\mathrm{dn}[2\kappa
(z-L)/m;m]}{2\mathrm{dn}[2\kappa (z-L)/m;m]},  \label{eq12b}
\end{eqnarray}
where
\[
m=\frac{k}{\sqrt{1+k^{2}}}=\frac{1}{\sqrt{1+(\gamma C/4\kappa
)^{2}}}.
\]
The parameter $C$ can be found using the transcendental equation
\begin{equation}
A_{0}^{2}=C\frac{1+\mathrm{dn}[2\kappa L/m;m]}{2\mathrm{dn}[2\kappa
L/m;m]}. \label{eq13}
\end{equation}

Finally, one can define transmission and \textquotedblleft
reflection\textquotedblright\ coefficients for the nonlinear coupler
as
\begin{eqnarray}
\mathfrak{T}
&=&\frac{P_{1}(L)}{P_{1}(0)}=\frac{C}{A_{0}^{2}}=\frac{2\mathrm{
\ dn}[2\kappa L/m;m]}{1+\mathrm{dn}[2\kappa L/m;m]},  \label{eq14a} \\
\mathfrak{R}
&=&1-\frac{P_{1}(L)}{P_{1}(0)}=\frac{1-\mathrm{dn}[2\kappa
L/m;m]}{1+\mathrm{dn}[2\kappa L/m;m]}.  \label{eq14b}
\end{eqnarray}

\begin{figure}[ht]
\centering
\includegraphics[scale=0.6]{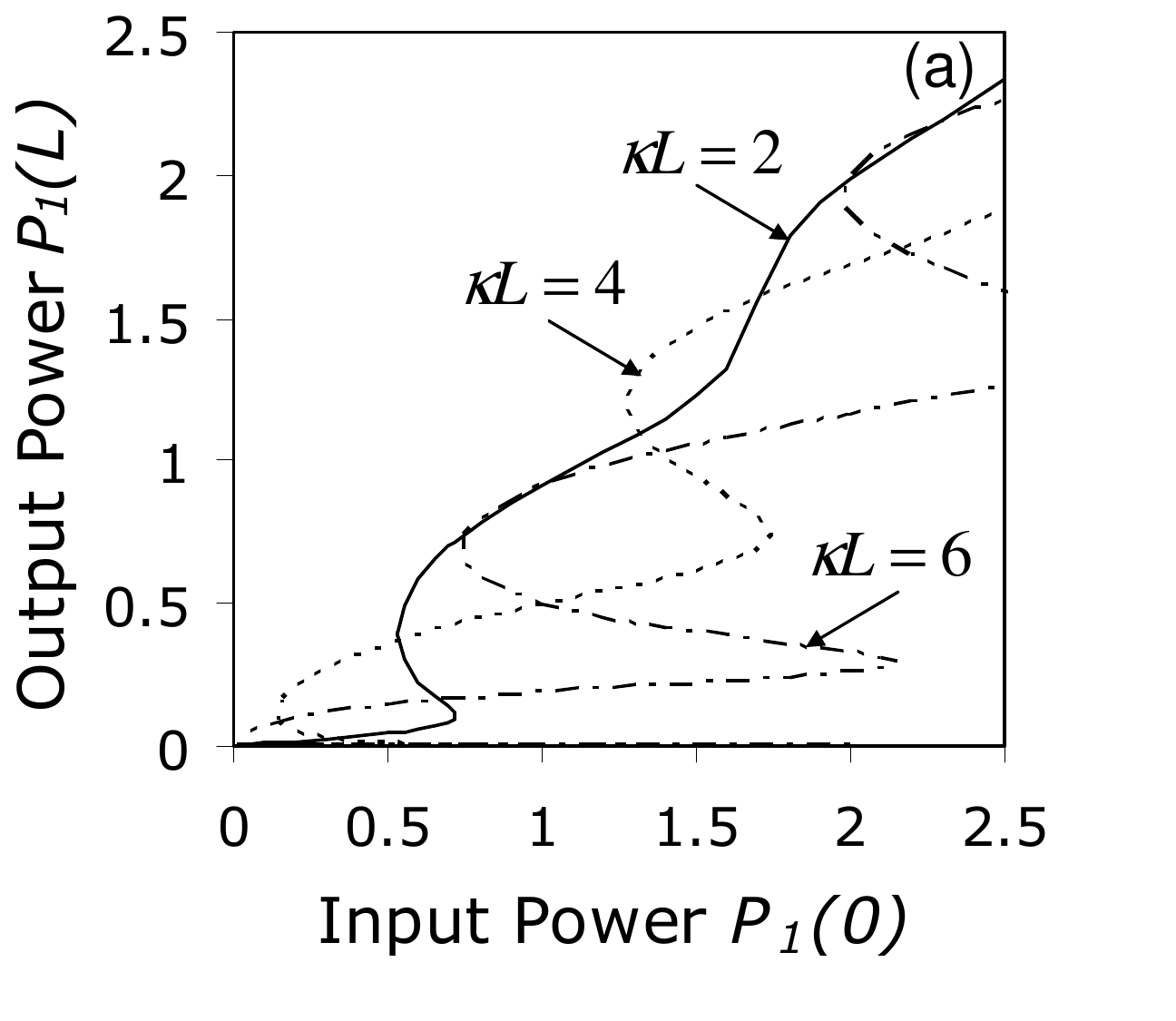}\includegraphics[scale=0.6]{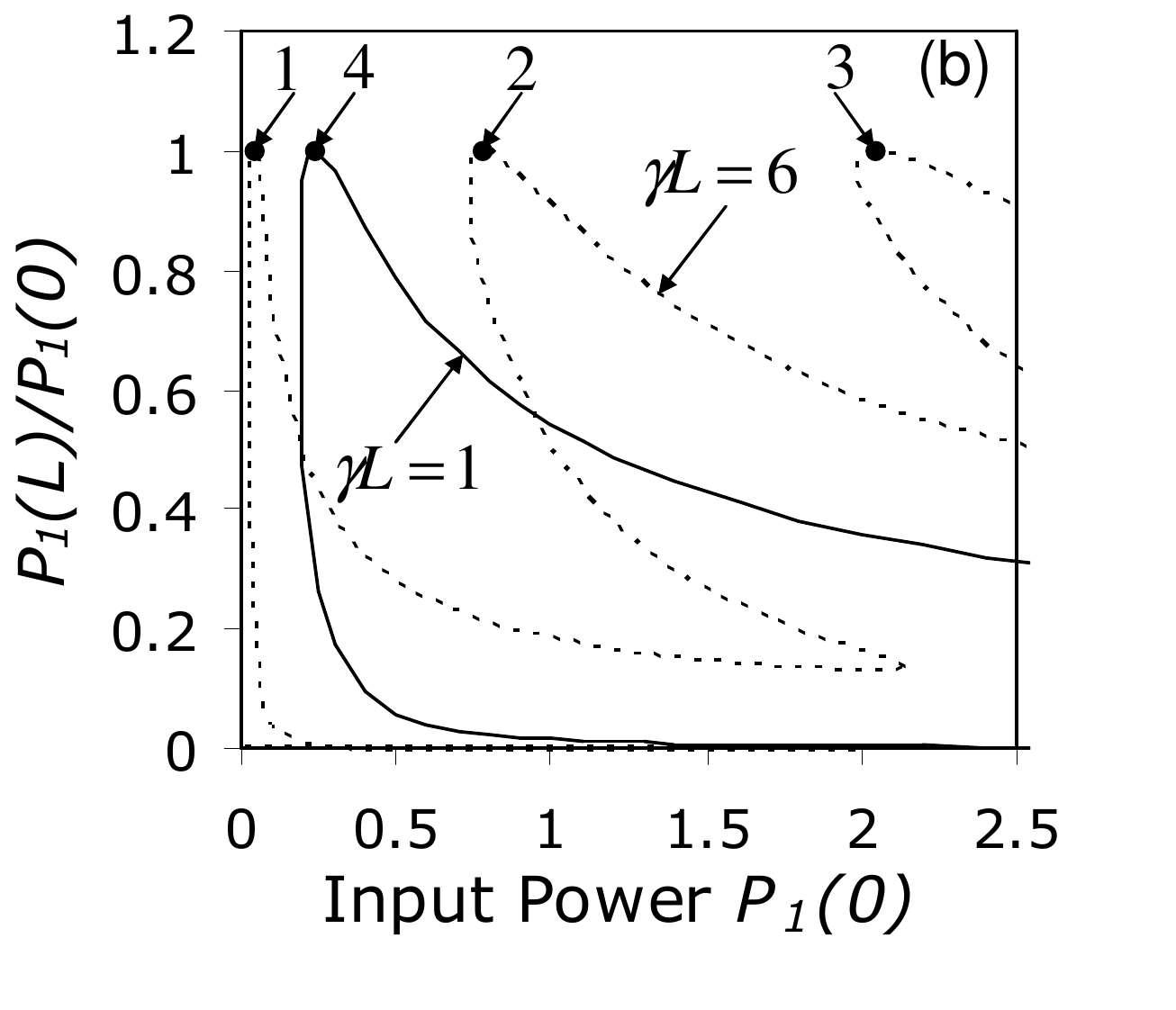}
\caption{\emph{(a) Output power $P_{1}(L)$  from channel 2 as a
function of input power $P_{1}(0)$  launched into channel 1 for
three values  of $kL=2$ (solid line), $kL=4$ (dashed line) and
$kL=6$ (dot-dashed line) when $\gamma L=6$. (b) Transmission
coefficient~defined as~$P_{1}(L)/P_{1}(0)$ as a function
of~$P_{1}(0)$ for $\gamma L=1$~(solid line) and~$\gamma L=6$ (dashed
line) when $\kappa L=6$. Transmission resonances are indicated by
the numbers 1,2,3, and 4.}}\label{Fig3}
\end{figure}

Figure~\ref{Fig3}(a) shows output power $P_{1}(L)$ as a function of
input power $P_{1}(0)$ for three values of $\kappa L=2~$(solid
line), $\kappa L=4~$ (dot-dashed line), and $\kappa L=6$ (dashed
line),~assuming $\kappa $\ is varying, the coupler length $L=1$~(in
the units of length) is fixed and $ \gamma L=6$. As the coupling
between the channels increases and the effective feedback mechanism
establishes, PIM-NIM NLC becomes bistable or more generally
multistable such as in the case of $\kappa L=6$. Its transmission
characteristics are very similar to those of the DFB
structures~\cite{8}-\cite{11}~with an important fundamental
difference that bistability in the PIM-NIM coupler is facilitated by
the effective feedback mechanism originating from the  NIM's
intrinsic property. Figure~\ref{Fig3}(b) shows the transmission
coefficient $\mathfrak{T}$ as a function of input power~$ P_{1}(0)$
for two values of ~$\gamma L=1~$(solid line) and $\gamma L=6$\
(dashed line), assuming $\gamma $\ is varying, the coupler length
$L=1$ is fixed and $\kappa L=6$. As the nonlinearity coefficient
decreases the threshold of bistability shifts to the higher values
as expected.

\begin{figure}[ht]
\centering
\includegraphics[scale=0.9]{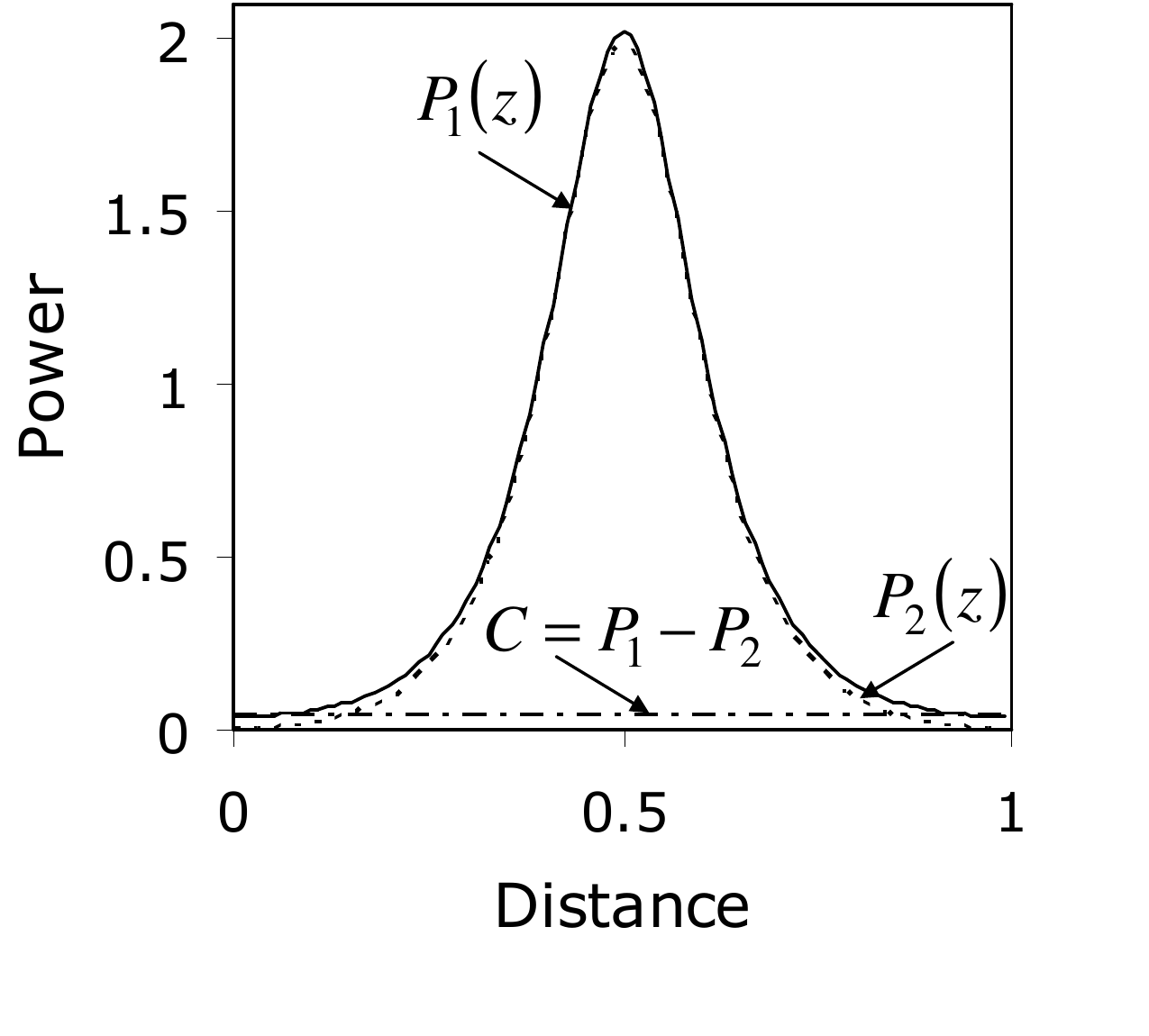}
\caption{\emph{Spatial power distribution in channel 1 (solid line),
in channel 2 (dashed line), and the constant of motion~\
$C=P_{1}-P_{2}$ (dot-dashed line) versus $z$ at transmission
resonance indicated by number `1' in
Fig.~\ref{Fig3}(b).}}\label{Fig4}
\end{figure}

The phenomenon of bistability in DFBs is closely related with the
notion of gap solitons~\cite{7}-\cite{15}. As shown in
Fig.~\ref{Fig3}(b), the transmission coefficient approaches
$\mathfrak{T}=1~$ at the points 1, 2, 3 and 4, suggesting the
existence of transmission resonances. At the resonance,
corresponding to a point 1 in Fig.~\ref{Fig3}(b), spatial power
distributions $P_{1}(z)$ (solid line) and~$P_{2}(z)$ (dashed line)
peak in the middle of the structure as shown in Fig.~\ref{Fig4}. The
dot-dashed line in Fig.~\ref{Fig4} shows the constant of the motion
$C=P_{1}-P_{2}$. At this transmission resonance incident light is
coupled to a soliton-like static entity that peaks in the middle of
the structure and is known as a gap soliton~\cite{10,11}.
Nonstationary gap solitons in PIM-NIM NLC analogous to those found
by Aceves and Wabnitz~\cite{13} in a context of periodic media will
be discussed elsewhere. It is notable that a gap soliton, usually
existing in periodic structures, forms in a uniform structure in the
case of the PIM-NIM coupler, owing to the new backward-coupling
mechanism in the PIM-NIM coupler.
In summary, we found that backward coupling between the modes
propagating in the PIM and NIM channels enabled by the basic
property of NIMs, oppositely directed phase and energy velocities,
results in optical bistability in PIM-NIM NLC and gap soliton
formation. These effects have no analogies in conventional PIM-PIM
couplers composed of uniform (homogeneous) waveguides with no
feedback mechanism.

\bigskip

This work was supported in part by the ARO Award \# 52730-PH,
ARO-MURI Award \# 50342-PH-MUR, by the NSF Award \# DMS-050989, by
State of Arizona grant TRIF (Proposition 301) and by the RFBR Award
\# 06-02-16406

\end{document}